\documentclass[a4paper,11pt]{article}
\usepackage{amsmath}
\usepackage{amsfonts}
\usepackage{amssymb}
\usepackage{bbm}
\usepackage{geometry}
\usepackage[parfill]{parskip} 
\usepackage{graphicx}
\usepackage{epstopdf}

\begin{document}

%% Making title, etc.
\title{{\bf Instantaneous Quantum Computation} }

\author{%
Dan~Shepherd\footnote{dan.shepherd@cesg.gsi.gov.uk, shepherd@compsci.bristol.ac.uk}~~and~Michael~J.~Bremner
\\
\small\it Department of Computer Science, University of
Bristol,\\ \small \it Woodland Road, Bristol, BS8 1UB, United Kingdom.
}

\date{December 13th, 2008}
\maketitle
%~~~~~~~~~~~~~~~~~~~~~~~~~~~~~~~~~~~~~~~~~~~~~~~~~~~~~~~~~~~~~~~~~~~~~~~~~~~~~~~~~~~~~~~~

%% Handy Dirac notation.
\def\ket#1{|\,#1\,\rangle}
\def\bra#1{\langle\, #1\,|}
\def\braket#1#2{\langle\, #1\,|\,#2\,\rangle}
\def\ketbra#1#2{\ket{#1}\bra{#2}}
\def\identity{\leavevmode\hbox{\small1\kern-3.8pt\normalsize1}}  %Makes an "open 1"
\def\span#1{\left< #1 \right>}

%% Formatting instructions
\def\lineacrosspage{\hbox to \hsize{\hfill\rule[5pt]{2.5cm}{0.5pt}\hfill}}
\def\FLAG{ \par \medskip \noindent \emph{[This Section Not Yet Complete]} \par \medskip }
\def\comment#1{}

%% A bit of maths.
\def\set#1{\{ #1\}}
\def\Prob#1{\mbox{Prob}(#1)}
\def\modulus#1{\left| #1 \right|}
\newcommand{\QED}{\nopagebreak\hspace*{\fill}\mbox{\rule[0pt]{1.5ex}{1.5ex}}}
\newcommand{\qed}{\mbox{\rule[0pt]{1.5ex}{1.5ex}}}
\newcommand{\half}{\mbox{$\textstyle \frac{1}{2}$} }
\def\indicator#1{\left\{ \phantom{\big|} #1 \phantom{\big|}\right\}}

\def\Z{\mathbbm{Z}}
\def\R{\mathbbm{R}}

%% Some environments
\newtheorem{theorem}{Theorem}
\newtheorem{definition}{Definition}
\newtheorem{lemma}{Lemma}
\newtheorem{example}{Example}
\newtheorem{property}{Property}
\newtheorem{proposition}{Proposition}
\newtheorem{corollary}{Corollary}
\newtheorem{conjecture}{Conjecture}
\def\proof{\textbf{Proof~:~}}

\def\Code{\mathcal{C}}
\def\C{\mathbbm{C}}
\def\E{\mathbbm{E}}
\def\F{\mathbbm{F}}
\def\P{\mathbbm{P}}
\def\a{\mathbf{a}}
\def\b{\mathbf{b}}
\def\c{\mathbf{c}}
\def\d{\mathbf{d}}
\def\e{\mathbf{e}}
\def\k{\mathbf{k}}
\def\l{\mathbf{l}}
\def\p{\mathbf{p}}
\def\q{\mathbf{q}}
\def\s{\mathbf{s}}
\def\w{\mathbf{w}}
\def\x{\mathbf{x}}
\def\X{\mathbf{X}}
\def\y{\mathbf{y}}
\def\Y{\mathbf{Y}}
\def\z{\mathbf{z}}
\def\zero{\mathbf{0}}
\def\one{\mathbf{1}}
\def\S{\mathbbm{S}}
\def\B{\mathcal{B}}
\def\eps{\varepsilon}

\def\eq#1{=_{\phantom|_{\!\!#1}}}  % This puts a tiny subscript after the = sign.
\def\pl#1{+_{\phantom|_{\!\!#1}}}
\def\mi#1{-_{\phantom|_{\!\!#1}}}
\def\om#1{\omega_{\phantom|_{\!\!#1}}}

\vspace{-0.3 cm}

%% ABSTRACT

\begin{abstract}

We examine theoretic architectures and an abstract model for a restricted class of quantum computation, called here \emph{temporally unstructured (``instantaneous'') quantum computation} because it allows for essentially no temporal structure within the quantum dynamics.
Using the theory of binary matroids, we argue that the paradigm is rich enough to enable sampling from probability distributions that cannot, classically, be sampled from efficiently and accurately.  
This paradigm also admits simple interactive proof games that may convince a skeptic of the existence of truly quantum effects.  
Furthermore, these effects can be created using significantly fewer qubits than are required for running Shor's Algorithm.  

\end{abstract}

%% SECTION 1

\def\IQP{\mathbf{IQP}}
\def\BPP{\mathbf{BPP}}
\def\BQP{\mathbf{BQP}}

\section{Introduction}

\subsection{Mathematical motivation}

It has often been said that underlying the power of quantum computers is the close connection between the computational model and the way we represent dynamics in quantum systems. This connection is implicit in the standard circuit model, where we require a universal gate set for an $n$-logical-qubit processor to be capable of simulating the dynamics of the $n$-qubit unitary group $SU(2^n)$.  While there are many equivalent models of (universal) quantum computing, and not all of them explicitly `generate' the special unitary group on $n$ qubits, they each simulate (to within some pre-defined precision) operations drawn from \emph{some} non-abelian unitary group on a set of qubits.  Our approach in this paper departs from this well trod path, by focussing almost exclusively on an abelian subgroup of the unitary group.  This approach is much more restrictive in the kinds of computation allowed, leading to a computational paradigm that lies somewhere between classical and universal quantum computing.

The non-abelian nature of quantum circuit elements is undoubtedly a crucial feature of universal quantum computing; for example, it imposes a clear physical limitation to the time-ordering of the gates in a circuit. In the standard model of quantum computation, the only circuits that can be performed in a single ``time-step'' are those composed only of single-qubit gates and two-qubit gates that act on disjoint sets of qubits. We often refer to such circuits as depth-1 circuits. 
When an abelian group is being used for the gates within a circuit, that circuit need not be depth-1 in the sense just described, though it will nonetheless be essentially devoid of temporal structure, since the order of the gates is immaterial.
Physically, the quantum circuit model can be interpreted as applying a controlled sequence of unitary operations, which can in turn be thought of as a sequence of Hamiltonian evolutions.  If any two consecutive gates in a sequence commute with one another, then their order in the sequence can be freely interchanged, or equivalently, their Hamiltonians can simply be combined additively, which corresponds to simultaneous evolution.  When \emph{all} gates commute, a single simultaneous Hamiltonian evolution describes the dynamics, whose terms are the individual gates.

\subsection{Physical motivation}

How can we tell when we have successfully built a quantum computer? Given that tomography quickly becomes difficult as the number of qubits in a system grows, it is pertinent to ask if there is a simpler way of verifying the success of a quantum computation. One way, which has already been attempted in several experiments, \emph{e.g.} \cite{lit:Van01, lit:Lan07, lit:Lu07, lit:Tame06}, would be to use the prototype quantum computer to find the solution to a problem which we think is difficult to solve on a classical computer.

For instance, the following scenario is generic.  Alice is a skeptic, she doesn't believe that Bob has a quantum device at his disposal.  Fortunately, she is relatively certain that classical computers can't efficiently find the prime factors of a large integer, whereas quantum computers can \cite{lit:Sh94} (although many qubits may be required for a convincing demonstration).  So she issues a challenge to Bob~: she chooses a large number for which she cannot find the prime factors and sends it to Bob. If Bob then sends back the prime factors of her number within a reasonable time period, she can easily convince herself that Bob must have had a quantum device at his disposal.

This scenario in particular is one which has been used in attempts to verify the success of several small-scale quantum computers \cite{lit:Van01, lit:Lan07, lit:Lu07}, though of course the numbers used were too small to be considered hard to factor classically. Unfortunately, so far as we know, Shor's factoring algorithm is a relatively difficult quantum algorithm to perform.  It is well known that it can be implemented in a circuit model using polynomial circuit depth and linear circuit width, or logarithmic depth with a larger width \cite{lit:CW00}, or even with constant depth if arbitrarily wide `fanout' gates are allowed \cite{lit:Hoy02}.  In either case, we'd apparently require a fully universal set of quantum gates, and more than a thousand logical qubits, for a convincing demonstration.  

In this paper, we use abelian dynamics to suggest a two-party protocol (with classical message-passing), which could be used to test remotely a quantum device, which we believe is physically far less complex than factorization.  
We conjecture that it is classically infeasible to simulate the quantum process in our protocol, and that our protocol is simpler to implement than all known versions of Shor's algorithm, not requiring anything like a universal gate set.

\subsection{Guide to the paper}

We introduce the paradigm $\IQP$, which stands for ``Instantaneous Quantum Polytime''.  It is a restricted model of quantum computation, which can also be thought of in terms of an oracle for computation.  Here `polytime' means that the process is bound to consume at most a polynomial amount of resource in any reasonable model of quantum computation, while `instantaneous' means that the algorithmic process itself is to contain no inherent temporal structure.
We give formal definitions, and argue that there are non-trivial applications for $\IQP$.  In particular, a two-party interactive protocol game is described and discussed.

These are our main points, to bear in mind throughout the paper~:
\begin{itemize}
  \item
We define a restricted model of quantum computation, called $\IQP$ (section \ref{sect:defs}).
  \item
We present several different quantum architectures (section~\ref{sect:architectures}) that can render computations in the $\IQP$ model, suggesting that it is a `lowest common denominator' of some `natural' ideas for computing based on `abelian' notions.
 \item
Although we know of no specific architecture where the Hamiltonians and measurements involved in the $\IQP$ paradigm really are genuinely `easy' to implement, nonetheless there is a clear sense in which the mathematics that underlies the computation is `easy'.  Perhaps the ``Graph State'' architecture (section~\ref{sect:architectures}) gives the clearest example of a practical computing idea.
  \item
We argue and conjecture that the probability distributions generated in the $\IQP$ paradigm (section \ref{sect:defs}) are not only classically hard to sample from approximately, but that there are actual polynomial-time protocols (section~\ref{sect:protocol}) that can be completed using an $\IQP$ oracle that (we believe) can't be completed classically in polynomial time.
 \item
We provide an explicit example of such a protocol, involving two parties.  The purpose of the protocol is simply for one party to prove to the other that they are capable of approximating a multi-qubit output distribution having characteristics that match the output distribution of a particular $\IQP$ process.  In the protocol, Alice designs a problem with a `hidden' property, and sends it to Bob;  Bob runs the problem through his $\IQP$ oracle several times, and sends the classical outputs back to Alice;  Alice then uses the secret `hidden' property to assess whether there is good evidence for Bob having used a real $\IQP$ oracle.  This is all done in section~\ref{sect:protocol}.
  \item
We make a pragmatic analysis of our suggested protocol, showing how to fine-tune its parameters in order to make plausible the conjecture that it really can't be `faked' classically (section~\ref{sect:heuristics}).
  \item
By analogy, this protocol is to quantum computation what Bell experiments are to quantum communication~: the simplest known `proof' of a distinctly quantum phenomenon.  (Of course, since there is no mathematical proof published to date of a separation between the power of quantum computation and classical computation, we still have to rely on certain computational hardness conjectures.)
  \item
Despite the existence of protocols apparently requiring an $\IQP$ oracle, we are unable to find any \emph{decision language} in $\BPP^\IQP$ that is not in $\BPP$.  There seems to be a sense in which the paradigm isn't able to `compute new information' (section~\ref{sect:heuristics}).
  \item
If there should one day be architectures that can implement $\IQP$ oracles---even though full-blown universal quantum computing remain an unresolved engineering challenge---then our protocol may be an important demonstrator of the power of quantum mechanics for quantum computing. 
\end{itemize}

Much of the mathematics used depends significantly on the theory of binary matroids and binary linear codes, and so we spend some time in section~\ref{sect:defs} recalling some basic definitions.  Readers interested in the main construction (the interactive game) should start at section~\ref{sect:protocol} and dip back into the earlier definitions where needed.  Pure mathematicians might prefer to start at section~\ref{sect:mathythingo};  cryptographers might particularly appreciate section~\ref{sect:heuristics};  whereas physicists may prefer section~\ref{sect:architectures}.

\section{The $\IQP$ paradigm}  \label{sect:defs}

In this section, we define what we call the ``X-program'' architecture, and use it to define the notion of $\IQP$ oracle.  The rest of the paper depends heavily on these definitions.  Note that the X-program architecture is not particularly `physical', but is easier to work with than the more physically relevant architectures discussed in section~\ref{sect:architectures}.

\subsection{X-programs}  \label{sect:Xprogs}

Recall that a Pauli $X$ operator acts on a single qubit, exchanging $\ket0$ with $\ket1$, \emph{i.e.} $X = \ketbra01 + \ketbra10$.  One can also think of $X$ as a Hamiltonian term, since $X \propto \exp( i\frac\pi2 X )$.  An X-program is essentially a Hamiltonian that is a sum of products of $X$s acting on different qubits.

In this architecture, allow for a set of $n$ qubits, initialised into the pure separable computational basis state $\ket{\zero}$.
The \emph{X-program} is specified as a (polysize) list of pairs $(\theta_\p, \p) \in [0,2\pi] \times \F_2^n$, so $\theta_\p$ is an angle and $\p$ is a string of $n$ bits.
Each such program element (pair) is interpreted as the action of a Hamiltonian on the qubits indicated by $\p$, applied for action\footnotemark{} $\theta_\p$~:
the Hamiltonian to apply is made up from a product of Pauli $X$ operators on the indicated qubits, and naturally these all commute.
\footnotetext{action = the integral of energy over time.}  
% $\int E~dt$}  
% energy $\times$ time}
%
This means that---in principle---the program elements could all be applied simultaneously~: their time ordering is irrelevant.
The measurement to be performed, once all the Hamiltonians have been applied, is simply a computational-basis measurement, and the program \emph{output} is simply that measurement result, regarded as a (probabilistic) sample from the vectorspace $\F_2^n$.

Combining this together, we see that the probability distribution for such an output is 
\begin{equation}  \label{eqn:dist1}
  \P(\X=\x) ~=~ \left| \bra\x ~\exp\left(~\sum_\p i\theta_\p \bigotimes_{j:p_j=1} X_j~\right)~ \ket{\zero^n} \right|^2.
\end{equation}

The \emph{output string} here is labelled $\x$.  The random variable $\X$ here (and throughout) codifies this probability distribution of classical output samples.

For almost all of our purposes, we are only interested in the case where the $\theta_\p$ action values are the same for every term.  When that condition applies, and the value $\theta$ is specified, the entire X-program can be represented using a $poly(n)$-by-$n$ binary matrix, with each row corresponding to a term in the Hamiltonian.
For example, the $7$-by-$4$ binary matrix
\begin{eqnarray}  \label{eqn:exampleP}
  P &=&  \left( \begin{array}{cccc}
1& 0& 0& 0 \\
1& 1& 0& 0 \\
0& 1& 1& 0 \\
1& 0& 1& 1 \\
0& 1& 0& 1 \\
0& 0& 1& 0 \\
0& 0& 0& 1
  \end{array}  \right) 
\end{eqnarray}
would represent the $4$-qubit Hamiltonian
\begin{eqnarray}
  H_{P,\theta} &=& 
  \theta \cdot ( X_1 + X_1X_2 + X_2X_3 + X_1X_3X_4 + X_2X_4 + X_3 + X_4 ).
\end{eqnarray}

\subsection{$\IQP$ oracle}

\begin{definition}
On input the explicit description of an X-program, we define an $\IQP$ oracle to be any computational method that efficiently returns a sample string from the probability distribution as given at line~(\ref{eqn:dist1}).
\end{definition}
\medskip

For a formal definition of what is meant by an $\IQP$ oracle, let it be any device that interfaces to a probabilistic Turing machine via an `oracle tape', so that if the oracle tape holds a description of a particular X-program ($P,\theta$ in the `constant action' case) at the time when the Turing machine calls its `implement oracle' instruction, then in unit time (or perhaps in time polynomial in the length of the description of $P$, \emph{i.e.} polynomial in $n$), a bitstring sample in $\F_2^n$ from the probability distribution at line~(\ref{eqn:dist1}) is created and written to the oracle tape, and control is passed back to the Turing machine to continue processing.

We write the \emph{overall paradigm}---of classical computation augmented by this oracle---as $\mathbf{BPP^{IQP}}$, to denote the fact that classical randomised polytime pre- and post-processing is usually to be considered allowed in a simulation, and to denote the fact that we don't much care which of several quantum architectures might be being used to supply the `\textbf{IQP}-power' of sampling from probability distributions of the form at line~(\ref{eqn:dist1}).
This notation is not necessarily supposed to indicate a particular class of \emph{decision languages} as such, but rather a particular class of computations.
The interactive proof games in section~\ref{sect:protocol} require the Prover to have access to an $\IQP$ oracle, and to access it a polynomial number of times, though these calls may be made in parallel and without precomputation.

\medskip
\begin{lemma}  \label{lem:thing}
The probability distribution given at line~(\ref{eqn:dist1}) is equivalent to the one given below.
\begin{eqnarray}  \label{eqn:distpaths}
  \P(\X=\x) &=& \left|~  \sum_{ \a ~:~ \a \cdot P = \x} 
                      ~~\prod_{\p ~:~ a_\p = 0}   \cos \theta_\p 
                        \prod_{\p ~:~ a_\p = 1} i \sin \theta_\p ~\right|^2.
\end{eqnarray}
\end{lemma}
\medskip

\proof
Let $P$ denote the $k$-by-$n$ binary matrix whose rows are the $\p$ vectors of the X-program under consideration.
Then using the fact that the Hamiltonian terms in an X-program all commute, we can think of the quantum amplitudes arising in an X-program implementation as a sum over paths, 
\begin{eqnarray}
  \lefteqn{ \bra{\x}~ \prod_\p 
            \left( \cos \theta_\p ~+~ i \sin \theta_p \bigotimes_{j:p_j=1} X_j \right) 
            ~\ket{\zero^n} }  \nonumber \\
  &=&
  \bra{\x} ~\sum_{ \a \in \F_2^k} 
    ~\prod_{\p ~:~ a_\p = 0}   \cos \theta_\p 
     \prod_{\p ~:~ a_\p = 1} i \sin \theta_\p 
    ~\prod_{j=1}^n X_j^{(\a \cdot P)_j} ~~\ket{\zero^n},
\end{eqnarray}
and hence derive a new form for the probability distribution accordingly.
\hfill \qed

\subsection{Binary matroids and Linear binary codes}

Before proceeding to the main topics of the paper, it behooves us to establish the link that these formulas have with the (closely related) theories of binary matroids and binary linear codes. 

\medskip
\begin{definition}
A linear binary code, $\Code$, of length $k$ is a (linear) subspace of the vectorspace $\F_2^k$, represented explicitly.
The elements of $\Code$ are called \emph{codewords}, and the Hamming weight $wt(c) \in [0..k]$ of some $c \in \Code$ is defined to be the number of 1s it has.  The rank of $\Code$ is its rank as a vectorspace.
\end{definition}
\medskip

Linear binary codes are frequently presented using \emph{generator matrices}, where the columns of the generator matrix form a basis for the code.  If $P$ is a generator matrix for a rank $r$ code $\Code$, then $P$ has $r$ columns and the codewords are $\{ P \cdot \d^T ~:~ \d \in \F_2^r \}$.

There are many different, isomorphic, definitions for matroids, \cite{lit:matroids}.  
We shall adopt the following definition.

\medskip
\begin{definition}
A $k$-point binary matroid is an equivalence class of matrices defined over $\F_2$, where each matrix in the equivalence class has exactly $k$ rows, and two matrices are equivalent (written $M_1 \sim M_2$) when for some ($k$-by-$k$) permutation matrix $Q$, the column-echelon reduced form of $M_1$ is the same as the column-echelon reduced form of $Q \cdot M_2$.  Here we take column-echelon reduction to delete empty columns, so that the result is full-rank.  Hence the rank of a matroid is the rank of any of its representatives.  
\end{definition}
\medskip

Less formally, this means that a binary matroid is like a matrix over $\F_2$ that doesn't notice if you rearrange its rows, if you add one of its columns into another (modulo 2), or if you duplicate one of its columns.  This means that a matroid is like the generator matrix for a linear binary code, but it doesn't mind if it contains redundancy in its spanning set (\emph{i.e.} has more columns than its rank) and it doesn't care about the actual order of the zeroes and ones in the individual codewords.
To be clear, when thinking of a matrix such as $P$ in line~(\ref{eqn:exampleP}), we are simultaneously thinking of its \emph{columns} as the elements of a spanning set for a \emph{code}, and its \emph{rows} as the points of a corresponding \emph{matroid}.
Because one cannot express a matroid independently of a representation, we consistently conflate notation for the matrix $P$ with the matroid $P$ that it represents.
 
Perhaps the main structural feature of a binary matroid is its \emph{weight enumerator polynomial}.  

\medskip
\begin{definition}  \label{def:WEP}
If the $k$ rows of binary matrix $P$ establish the points of a $k$-point matroid, then the weight enumerator of the matroid is defined to be the weight enumerator of the $k$-long code $\Code$ spanned by the columns of $P$, which in turn is defined to be the bivariate polynomial
\begin{eqnarray}
  WEP_\Code(x,y)  &=&  \sum_{\c \in \Code} x^{wt(\c)} y^{k-wt(\c)}.
\end{eqnarray}
\end{definition}
\medskip

This is well-defined, because the effect of choosing a different matrix $P$ that represents the same binary matroid simply leads to an isomorphic code that has the same weight-enumerator polynomial as the original code $\Code$.

\subsection{Bias in probability distributions}

\begin{definition}  \label{def:bias}
If $\X$ is a random variable taking values in $\F_2^n$, and $\s$ is any element of $\F_2^n$, then the \emph{bias} of $\X$ in direction $\s$ is simply the probability that $\X \cdot \s^T$ is zero, \emph{i.e.} the probability of a sample being orthogonal to $\s$.
\end{definition}
\medskip
 
Let us now consider an X-program on $n$ qubits that has constant action value $\theta$, whose Hamiltonian terms are specified by the rows of matrix $P$, as discussed earlier.
Then we can use lemma~\ref{lem:thing} to obtain the following expression of bias, for any binary vector $\s \in \F_2^n$~:
\begin{eqnarray}  \label{eqn:walshpaths}
  \P(\X \cdot \s^T=0) 
    &=&  \sum_{\x ~:~ \x \cdot \s^T = 0} ~
         \left|~ \sum_{ \a ~:~ \a \cdot P = \x} ~
             (\cos \theta)^{k-wt(\a)}  (i \sin \theta)^{wt(\a)} 
        ~\right|^2.
\end{eqnarray}

Since it would obviously be nice to interpret this expression as the evaluation of a weight enumerator polynomial, we are led to define $P_\s$ to be the submatrix of $P$ obtained by deleting all rows $\p$ for which $\p \cdot \s^T = 0$, leaving only those rows for which $\p \cdot \s^T = 1$.  We call\footnotemark{} the number of rows remaining $n_\s$.  
\footnotetext{$n_\s$ is here being used for the length of the code $\Code_\s$ in deference to the usual practice of reserving the letter $n$ for code lengths.  This $n_\s$ is counting a number of rows, and should not be confused with the $n$ used earlier for counting a number of columns.}
This in turn leads to the code $\Code_\s$ being the span of the columns of $P_\s$, and likewise a submatroid\footnotemark{} is correspondingly defined.
\footnotetext{also called a \emph{matroid minor}}

\medskip
\begin{theorem}  \label{thm:bias}
  When considering constant-action X-programs, the bias expression $\P(\X \cdot \s^T=0)$ for the random variable $\X$ introduced at line~(\ref{eqn:dist1}) depends only on the action value $\theta$ and (the weight enumerator polynomial of) the $n_\s$-point matroid $P_\s$, as defined above. 
Moreover,  if $\Code_\s$ is a binary code representing the matroid $P_\s$, then the following formula\footnotemark{} expresses the bias~:
\begin{eqnarray}  \label{eqn:walshcode}
    \P(\X \cdot \s^T=0) 
    &=& \E_{\c \sim \Code_\s} \left[~ \cos^2\Bigl(~ \theta( n_\s ~-~ 2 \cdot wt(\c) ) ~\Bigr) ~\right].
\end{eqnarray}
\end{theorem}
\medskip
\footnotetext{Subscripts on expectation operators indicate a variable ranging uniformly over its natural domain.}

\proof
See the appendix for a proof.  \hfill \qed \medskip

To recap, this means that if we run an X-program using the action value $\theta$ for all program elements, then the probability of the returned sample being orthogonal to an $\s$ of our choosing (`orthogonal' in the $\F_2$ sense of having zero dot-product with $\s$) depends only on $\theta$ and on the (weight enumerator polynomial of the) linear code obtained by writing the program elements $\p$ as rows of a matrix and ignoring those that are orthogonal to $\s$.

There is a definition in the literature for \emph{weighted matroids}, which in this context would correspond to allowing different $\theta$ values for different terms in the Hamiltonian of an X-program.  While mathematically (and physically) natural, such considerations would not help with the clarity of our presentation.

We emphasise at this point the value of theorem~\ref{thm:bias}~: it means that for any direction $\s \in \F_2^n$, the bias of the output probability distribution from an X-program $(P,\theta)$ in the direction $\s$ depends \emph{only} on $\theta$ and the rows of $P$ that are \emph{not} orthogonal to $\s$, and not at all on the rows of $P$ that \emph{are} orthogonal to $\s$. 
Moreover, the bias in direction $\s$ depends \emph{only} on the \emph{matroid} $P_\s$, and not on the particular \emph{matrix} $P_\s$ that represents it.  That is, directional bias (definition~\ref{def:bias}) is a matroid invariant.

Note that whenever $A$ is an $n$-by-$n$ invertible matrix over $\F_2$, then 
  $\p \cdot \s^T  ~=~  \p \cdot A \cdot A^{-1} \cdot \s^T  ~=~  
                       (\p \cdot A) \cdot (\s \cdot A^{-T})^T$, 
so any invertible column operation on matrix $P$ accompanies an invertible change of basis for the set of directions of which $\s$ is a member.  Note also that appending or removing an all zero column to $P$ has the effect of including or excluding a qubit on which no unitary transformations are performed.
Thus if $P_\s$ is a submatroid of $P$ by point-deletion, as described earlier, then if the invertible column transformation $A$ is applied to the matrix $P$ that represents the matroid $P$, then the same \emph{matroid} that was formerly called $P_\s$ is still a submatroid, but now it is represented by the matrix $P_{\s \cdot A^{-T}}$.
Likewise, appending or removing a column of zeroes to $P$ necessitates an extra zero be appended or removed from any $\s$ that serves as a direction for indicating a submatroid.
This is purely an issue of representation, and we consider that intuition about these objects is aided by taking an `abstractist' approach to the geometry.

\subsection{Entropy, and trivial cases}  \label{sect:entropy}

Because it will be useful later, we will define the R\'enyi entropy (collision entropy) of a random variable, before exemplifying theorem~\ref{thm:bias} and proceeding with the main construction of the paper.

\medskip
\begin{definition}
The collision entropy, $S_2$, of a discrete random variable, $\X$, measures the randomness of the sampling process by measuring the likelihood of two (independent) samples being the same.  It is defined by 
\begin{eqnarray} \label{eqn:Renyi}
  2^{-S_2}  &=&  \sum_\x \P(\X=\x)^2
           ~~=~~  \E_\s \left[~ \Bigl(~ 2\P( \X \cdot \s^T = 0 ) - 1 ~\Bigr)^2 ~\right].
\end{eqnarray}
\end{definition}
\medskip

And so there are a few `easy cases' for our $\X$ random variable of lemma~\ref{lem:thing} that should be highlighted and dismissed up front~:

\medskip
\begin{lemma} 
For a constant-action X-program, if $\theta$ is\ldots

\begin{itemize}
  \item
\ldots a multiple of $\pi$, then the returned sample will always be $\zero$. 
The collision entropy will be zero.
  \item
\ldots an odd multiple of $\pi/2$, then the returned sample will always be $\sum_{\p \in P} \p$.  
The collision entropy is zero.  
  \item
\ldots an odd multiple of $\pi/4$, then the collision entropy need not be zero, but the probability distribution will be classically simulable to full precision.
\end{itemize}
\end{lemma}
\medskip

\proof
In the first case, considering line~(\ref{eqn:distpaths}), there is then a $\sin(\pi)=0$ factor in every term of the probability, except where $\x=\zero$.

In the second case, considering again line~(\ref{eqn:distpaths}), there is then a $\cos(\pi/2)=0$ factor in every term, except where all the $\p$ vectors are summed together to give $\x$.  The same can also be deduced from theorem~\ref{thm:bias}, which implies that $\x$ will be surely orthogonal to $\s$ exactly when $n_\s$ is even, \emph{i.e.} exactly when an even number of rows of $P$ are \emph{not} orthogonal to $\s$, \emph{i.e.} exactly when $\sum_{\p \in P} \p$ \emph{is} orthogonal to $\s$.

For the third case, if $\theta$ is an odd multiple of $\pi/4$, then all the gates in the program would be Clifford gates.  By the Gottesman-Knill theorem there is then a classically efficient method for sampling from the distribution, by tracking the evolution of the system using stabilisers, \emph{etc.}
\hfill \qed \medskip

For other sufficiently different values of the action parameter, classical intractibility becomes a plausible conjecture (\emph{cf.} \cite{lit:SWVC08, lit:SWVC2}). 
In particular, the remainder of this paper will specialise to the case $\theta = \pi/8$, since we are able to make all our points about the utility of $\IQP$ even with this restriction.

\section{Interactive protocol}  \label{sect:protocol}

One would naturally like to find some `use' for the ability to sample from the probability distribution that arises from a temporally unstructured quantum polytime computation; a `task' or `proof' that can be completed using \emph{e.g.} an X-program, which could not be completed by purely classical means.
In this section we develop our main construction; a two-player interactive protocol game in which a Prover uses an $\IQP$ oracle simply to demonstrate that he does have access to an $\IQP$ oracle.

\subsection{At a glance}

There are three aspects of design involved in specifying an actual ``Alice \& Bob'' game~:

\begin{itemize}
  \item[A)]   a code/matroid construction, for Alice to select a problem, to send to Bob,
  \item[B)]   an architecture or technique by which Bob to take samples from the $\IQP$ distribution of the challenge he receives, to send back to Alice,
  \item[A')]  an hypothesis test for Alice to use to verify (or reject) Bob's attempt.
\end{itemize}
\medskip

%In what follows we will address the first of these points.
We have already defined the X-program architecture, and in section~\ref{sect:architectures} we discuss some alternatives that Bob might like to try.
In section~\ref{sect:heuristics} we make an analysis of some classical cheating strategies for Bob, in case he can't lay his hands on a quantum computer.
The details of precisely how to make a good hypothesis test are omitted from this paper for the sake of brevity, but sourcecode is available on our website (see section~\ref{sect:challenge}).

Alice plays the role of the Challenger/Verifier, while Bob plays the role of the Prover.
Alice uses secret random data to obfuscate a `causal' matroid $P_\s$ inside a larger matroid $P$, and the latter she publishes (as a matrix) to Bob.  Bob interprets matrix $P$ as an X-program to be run several times, with $\theta = \pi/8$.  He collects the returned samples, and sends them to Alice.  Alice then uses her secret knowledge of `where' in $P$ the special $P_\s$ matroid is hidden, in order to run a statistical test on Bob's data, to validate or refute the notion that Bob has the ability to run X-programs.

This application is perhaps the simplest known protocol, requiring (say) $\sim 200$ qubits, that could be expected to convince a skeptic of the existence of some \emph{computational} quantum effect.  The reason for this is that there seems to be no classical method to fake even a \emph{classical transcript} of a run of the interactive game between Challenger and Prover, without actually \emph{being} (or subverting the secret random data of) the classical Challenger.

\subsection{Concept overview}

Consider therefore the following game, between Alice and Bob.
Alice, also called the Challenger/Verifier, is a classical player with access to a private random number generator.
Bob, also called the Prover, is a supposedly quantum player, whose goal is to convince Alice that he can access an $\IQP$ oracle, \emph{i.e.} run X-programs.  The rules of this game are that he has to convince her simply by sending classical data, and so in effect Bob offers to act as a remote $\IQP$ oracle for Alice, while Alice is initially skeptical of Bob's true $\IQP$ abilities.

\subsubsection{Alice's challenge}

The game begins with Alice choosing some code $\Code_\s$ that has certain properties amenable to her analysis. 
She chooses the code $\Code_\s$ in such a way that there is a $\theta$ for which the (quantum) expectation value at line~(\ref{eqn:walshcode}) of theorem~\ref{thm:bias} is somewhere well within $(\frac12, 1)$, and for which the corresponding expectation value that arises from the best-known classical approaches to `cheating' (\emph{e.g.} presumably the one at line~(\ref{eqn:classfromcode}) of section~\ref{sect:classical}, in case $\theta=\pi/8$) is significantly smaller.

She then finds a matrix $P_\s$ whose columns generate the code (not necessarily as a basis), and ensures that there is some $\s$ that is not orthogonal to any of the rows of $P_\s$.  The vector $\s$ should be thought of not as a structural property of the code $\Code_\s$, but as a `locator' that can be used to `pinpoint' $P_\s$ even after is has later been obfuscated. 

\emph{Obfuscation} of $P_\s$ is achieved by appending arbitrary rows that \emph{are} orthogonal to $\s$.  This gives rise to matrix $P$.  The matroid $P$ has $P_\s$ as a submatroid, in the sense that removal of the correct set of rows will recover $P_\s$.
Alice publishes to Bob a representation of matroid $P$ that hides the structure that she has embedded.  Random row permutations are appropriate, and reversible column operations likewise leave the matroid invariant (though the latter will affect $\s$ and must therefore be tracked by Alice).

\subsubsection{Bob's proof}

Bob, being $\mathbf{BPP^{IQP}}$-capable by hypothesis, may interpret the published $P$ as an X-program, to be run with the (constant) action set to $\theta = \pi/8$ (say).  He will be able to generate random vectors which independently have the correct bias in the (unknown to him) direction $\s$, \emph{i.e.} the correct probability of being orthogonal to Alice's secret $\s$, in accordance with theorem~\ref{thm:bias}.  Although he may still be entirely unable to recover this $\s$ from such samples, he nonetheless can send to Alice a list of these samples as proof that he is $\mathbf{BPP^{IQP}}$-capable.  

Note that Bob's strategy is error-tolerant, because if each run of the \textbf{IQP} algorithm were to use a `noisy' $\theta$ value, then the overall proof that he generates will still be valid, providing the noise is small and unbiased and independent between runs.  Note also that he can manage several runs in one oracle call, if desired, simply by concatenating the matrix $P$ with itself diagonally.  That is to say, we even avoid classical temporal structure (\emph{adaptive feed-forward}) on Bob's part.

\subsubsection{Alice's verification}

Since Alice knows the secret value $\s$, and can presumably compute the value $\P(\X \cdot \s^T=0)$ from the code's weight enumerator polynomial (see theorem~\ref{thm:bias} and recall that she is free to choose any $\Code_\s$ that suits her purpose), it is not hard for her to use a hypothesis test to confirm that the samples Bob sends are \emph{commensurate} with having been sampled independently from the same distribution that an X-program generates.
That is to say, Alice will not try to test whether Bob's data \emph{definitely fits the correct $\IQP$ distribution,} but she will ensure that it has the particular characteristic of a strong bias in the secret direction $\s$.  This enables her to test the null hypothesis that Bob is cheating, from the alternative hypothesis that Bob has non-trivial quantum computational power.  

\emph{This requires belief in several conjectures on Alice's part.}
She must believe that there is a classical separation between quantum and classical computing; in particular that $\IQP$ is not classically efficiently approximately simulable---at least she must believe that Bob doesn't know any good simulation tricks.
She must believe that her problem is hard---at least she should believe that the problem of identifying the location of $P_\s$ within $P$ is not a $\BPP$ problem---on the assumption that the matroid $P_\s$ is known.

If he passes her hypothesis test, Bob will have `proved' to Alice that he ran a quantum computation on her program, provided she is confident that there is no feasible way for Bob to simulate the `proof' data classically efficiently, \emph{i.e.} provided she has performed her hypothesis test correctly against a plausibly best null hypothesis.
In particular, Alice should ensure a large collision entropy for the true ($\mathbf{IQP}$) distribution, since she will want to remove all `short circuits' (\emph{i.e.} all the empty rows and all the duplicate rows) from Bob's data, before testing it, to make a test that is both fair and efficient.  Otherwise it would be too easy for Bob to generate a set of data that has a strong bias in very many directions simultaneously; and it would be tedious for Alice to confirm that he has not cheated in this way if she did not remove the short circuits.

\subsubsection{Significance}

This kind of interactive game could be of much significance to validation of early quantum computing architectures, since it gives rise to a simple way of `tomographically ascertaining' the actual presence of at least \emph{some} quantum computing, modulo some relatively basic complexity assumptions.  
In this sense it is to quantum computation what Bell violation experiments are to quantum communication.%
\footnote{We have serendipitously identified a construction for which the probability gap---quantum $85.4\%$ over classical $75\%$---precisely matches the gap available in Bell's inequalities.  See lemma~\ref{lem:QRCode}.} 

Of course, this protocol really comes into its own when the architecture being tested happens to have the undesirable engineering feature of being unable to sustain long-term quantum coherence, and therefore perhaps only ever being capable of shallow-depth computation.  Unfortunately, the prescription for X-programs given in section~\ref{sect:defs} requires Hamiltonian terms that act across potentially hundreds of qubits, and the alternative architectures discussed in section~\ref{sect:architectures} have similar physical drawbacks that still make this paradigm extremely challenging for today's engineering.

Note that this `testing concept' does not use the $\mathbf{IQP}$ paradigm to \emph{compute any data that is unknown to everyone}, nor does it directly provide Bob with any `secret' data that could be used as a witness to validate an $\mathbf{NP}$ language membership claim.  Its only effect is to provide Bob with data that he \emph{can't} use for any purpose other than to pass on to Alice as a `proof' of $\mathbf{IQP}$-capability.  It is an open problem to find something more commonly associated with computation---perhaps deciding a decision language, for example---that can be achieved specifically by the $\mathbf{BPP^{IQP}}$ paradigm.

\subsection{Recommended construction method}  \label{sect:recommend}

Here is a specific example of a construction methodology (with implicit test methodology) for Alice, which we conjecture to be asymptotically secure (against cheating Prover) and efficient (for both Prover and Verifier).

\subsubsection{Recipe for codes}

The family of codes that we suggest Alice should employ within the context of the game outlined above are the \emph{quadratic residue codes}.  These will be shown to have the significant property that there is a non-negligible gap between the quantum- and best-known-classical-approximation expectation values for the bias in the secret direction, both of which are significantly below 1.  (The bias for a truly quantum-enabled Bob has already been defined in theorem~\ref{thm:bias}, in terms of the weight-enumerator polynomial of the causal code.  For a classically cheating Bob, we discuss the best-known classical strategies and their biases in section~\ref{sect:classical}.)

Consider a quadratic residue code over $\F_2$ with respect to the prime $q$, chosen so that $q+1$ is a multiple of eight.  The rank of such a code is $(q+1)/2$, and the length is $q$.
A quadratic residue code is a cyclic code, and can be specified by a single cyclic generator.  There are several ways of defining these, but the simplest definition is to take the codeword that has a 1 in the $j$th place if and only if the Legendre symbol\footnotemark{} $\left(\frac{j}{q}\right)$ equals 1, \emph{i.e.} if and only if $j$ is a non-zero quadratic residue modulo $q$.
\footnotetext{This Legendre symbol is equivalent, modulo $q$, to $j^{(q-1)/2}$.}
For example, if $q=7$ (the smallest example) then the non-zero quadratic residues modulo $q$ are $\{1,2,4\}$, and so the quadratic residue code in question is the rank-$4$ code spanned by the various rotations of the generator $(0,1,1,0,1,0,0,0)^T$.  A basis for this code is found in the columns of the matrix at line~(\ref{eqn:exampleP}).

\medskip
\begin{lemma}  \label{lem:QRCode}
When $q$ is a prime and 8 divides $q+1$, then there is a unique quadratic residue code $\Code$ (up to isomorphism) of length $q$ over $\F_2$, having rank $(q+1)/2$, and it satisfies
\begin{eqnarray}  \label{eqn:QRCodestats1}
  \E_{\c \sim \Code} \left[~ \cos^2\Bigl(~ \frac\pi8 ( q ~-~ 2 \cdot wt(\c) ) ~\Bigr) ~\right]
  &=& \cos^2( \pi/8 ) ~~=~~ 0.854\ldots
\end{eqnarray}
Moreover, it also satisfies 
\begin{eqnarray}  \label{eqn:QRCodestats2}
  \P\Bigl(~ \c_1^T \cdot \c_2 = 0 ~|~ \c_1, \c_2 \sim \Code ~\Bigr)
  &=& 3/4 ~~=~~ 0.75,
\end{eqnarray}
which is relevant to certain classical strategies (section~\ref{sect:classical}).
\end{lemma}
\medskip

\proof
The proof of the \emph{rank} of the code and its uniqueness are well established results from classical coding theory \cite{lit:macwilliams}. Other classical results of coding theory include that quadratic residue codes are a parity-bit short of being self-dual and doubly even.  That is, the extended quadratic code, with length $q+1$, obtained by appending a single parity bit to each codeword, has every codeword weight a multiple of 4 and every two codewords orthogonal.

For line~(\ref{eqn:QRCodestats1}) this means that the (unextended) code has codeword weights which, modulo 4, are half the time 0 and half the time $-1$.  On putting these values into the left side of the formula, we immediately obtain the right side.  
For line~(\ref{eqn:QRCodestats2}) this means that in the (unextended) code, any two codewords are non-orthogonal if and only if they are both odd-parity, which happens a quarter of the time~: from which the formula follows.  \hfill \qed \medskip

The corollary here is that if Alice uses one of these codes for her `causal' $\Code_\s$, then if Bob runs a series of X-programs (with constant $\theta = \pi/8$) described by the (larger) matrix $P$, the data samples he recovers should be orthogonal to the hidden $\s$ about $85.4\%$ of the time (\emph{cf}. theorem~\ref{thm:bias}); whereas if Bob tries to cheat using the classical strategy outlined in section~\ref{sect:heuristics}, then his data samples will tend to be orthogonal to the hidden $\s$ only about $75\%$ of the time (\emph{cf}. lemma~\ref{lem:classprob}).  
Alice's hypothesis test therefore basically consists in measuring this single characteristic, after having filtered duplicate and null data samples from Bob's dataset.
We would conjecture that Bob has no pragmatic way of boosting these signals, at least not without feedback from Alice, or exponential resources.

Note that \emph{with} exponential time on his hands, Bob could choose to simulate classically an $\IQP$ oracle, in order to obtain a dataset with a bias in direction $\s$ that is approximately $85.4\%$.  Alternatively, he could consider every possible $\s$ in turn, and test to see whether the matroid obtained by deleting rows orthogonal to his guess is in fact correspondent to a quadratic residue code, assuming he knew that this had been Alice's strategy.

\subsubsection{Recipe for obfuscation}

Having chosen $q$ as outlined above, and constructed a $q$-by-$(q+1)/2$ binary matrix generating a quadratic residue code, Alice needs to obfuscate it.  The easiest way to manage this process is not to start with a particular secret $\s$ in mind, but rather to recognise the obfuscation problem as a \emph{matroid} problem, proceeding as follows~:
\begin{itemize}
  \item 
Append a column of 1s to the matrix~: this does not change the code spanned by its columns since the all-ones (full-weight) vector is always a codeword of a quadratic code.
Other redundant column codewords may also be appended, if desired.  
  \item
Append many (say $q$) extra rows to the matrix, each of which is random, subject to having a zero in the column lately appended.  This gives rise to a $2q$-point matroid, and  ensures that there now \emph{is} an $\s$ such that the causal submatroid (quadratic residue matroid) is defined by non-orthogonality of the rows to that $\s$.
  \item
Reorder the rows randomly.  This has no effect on the matroid that the matrix represents, nor on the hidden causal submatroid.  Nor does it affect $\s$, the `direction' in which the sumbatroid is hidden.
  \item
Now column-reduce the matrix.  There is no (desirable) structure within the particular form of the matrix before column-reduction, nothing that affects either codes or matroids.  Echelon-reduction provides a canonical representative for the overall matroid, while stripping away any redundant columns that would otherwise cost an unnecessary qubit, when interpreted as an X-program.  By providing a canonical representative, it closes down the possibility that information in Alice's original construction of a basis for her causal code might leak through to Bob, which might be useful to him in guessing $\s$.  Rather more importantly, this reduction actually serves to \emph{hide} $\s$.  (We can be sure by zero-knowledge reasoning that this hiding process is random~: echelon reduction is canonical and therefore supervenes any column-scrambling process, including a random one.)
  \item
Finally, one might sort the rows, though this is unnecessary.  The resulting matrix is the one to publish.  It will have at least $(q+1)/2$ columns, since that is the rank of the causal submatroid hidden inside.  
\end{itemize}

\subsubsection{Mathematical problem description} \label{sect:mathythingo}

This method of obfuscation amounts to---mathematically speaking---a situation whereby for each suitable prime $q$, we start by acknowledging a particular (public) $q$-point binary matroid $Q$, \emph{viz} the one obtained from the QR-Code of length $q$.  Then an `instance' of the obfuscation consists of a published $2q$-point (say) binary matroid $P$, and there is to be a hidden ``obfuscation'' subset $O$ such that $Q = P\backslash O$;
and the practical instances occur with $P$ chosen effectively at random, subject to these constraints.  (One could choose to make $O$ bigger than $q$ points if that were desired.)  This has the feel of a fairly generic hidden substructure problem, so it seems likely that it should be \textbf{NP}-hard to determine the location of the hidden $Q$, given $P$ and the appropriate promise of $Q$'s existence within.  
More syntactically, we should like to prove that it is \textbf{NP}-complete to decide the related matter of \emph{whether or not} $P$ is of the specified form, given only a matrix for $P$.  Clearly this problem is in \textbf{NP}, since one could provide $Q$ \emph{in the appropriate basis} as witness.  We conjecture this problem to be \textbf{NP}-complete.
 
\medskip
\begin{conjecture}  \label{conj:NPc}
The language of matroids $P$ that contain a quadratic-residue code submatroid $Q$ \emph{by point deletion}, where the size of $Q$ is at least half the size of $P$, is \textbf{NP}-complete under polytime reductions. 
\end{conjecture}
\medskip

These sorts of conjecture are apparently independent of conjectures about hardness of classical efficient $\IQP$ simulation, since they indicate that \emph{actually identifying the hidden data} is hard, even for a universal quantum computer.
Even should this conjecture prove false, we know of no reason to think that a quantum computer would be much better than a classical one at finding the hidden $Q$, notwithstanding Grover's quadratic speed-up for exhaustive search. 

One might compare the structure of conjecture~\ref{conj:NPc} to that of the following important \emph{theorem} from graph theory~:
\begin{proposition}
The language of graphs $G$ that contain a complete graph $K$ \emph{by vertex deletion}, where the size of $K$ is at least half the size of $G$, is \textbf{NP}-complete under polytime reductions. 
\end{proposition}

This is a classic result, see \emph{e.g.} \cite{lit:Papa}, where the problem in different guises is called `Clique' and `Independent Set' and `Node Cover'.

\subsection{Challenge}  \label{sect:challenge}

It seems reasonable to conjecture that, using the methodology described, with a QR-code having a value $q \sim 500$, it is very easy to create randomised Interactive Game challenges for $\mathbf{BPP^{IQP}}$-capability, whose distributions have large entropy, which should lead to datasets that would be easy to validate and yet infeasible to forge without an $\IQP$-capable computing device (or knowledge of the secret $\s$ vector).  We propose such challenges as being appropriate `targets' for early quantum architectures, since such challenges would essentially seem to be the simplest ones available (at least in terms of inherent temporal structure and number of qubits) that can't apparently be classically met.  

Accordingly, we have posted on the internet (http://quantumchallenges.wordpress.com) a \$25 challenge problem, of size $q=487$, to help motivate further study.
This challenge website includes the source code (C) used to make the challenge matrix, and also the source code of the program that we will use to check candidate solutions, excluding only the secret seed value that we used to randomise the problem.

\section{Heuristics}  \label{sect:heuristics}

Next we address in more detail the reasons for thinking the problem classically intractable, and also give an accounting of our failure to find a \emph{decision language} for proving the worth of $\IQP$.

\subsection{Hardness of strong simulation}

In support of the supposed complexity of this paradigm, Terhal and Divincenzo \cite{lit:TD02}, and Aaronson \cite{lit:Aa04}, have already showed that it is \textbf{PP-complete} to \emph{strongly simulate} the generic probability distributions that arise hence.  

\medskip
\begin{lemma}
  It is $\mathbf{P^{GapP}}$-hard to determine the numerical value of $\P(\X=\zero)$ (as defined in section~\ref{sect:defs}, line~(\ref{eqn:dist1})) to within exponential precision, for arbitrary matroids.
\end{lemma}
\medskip

\proof
Let $Ker_L(P) = \{ \a^T : \a \cdot P = \zero \}$ denote the linear code for which $P$ is a parity-check matrix, and note from line~(\ref{eqn:distpaths}) and definition~\ref{def:WEP} that the probability in question is a function of the weight-enumerator polynomial of this code.  Specifically,
\begin{eqnarray}  
  \P(\X=\zero) 
    &=&  \left|~ WEP_{Ker_L(P)}(~ \cos \theta, ~i \sin \theta ~) ~\right|^2.
\end{eqnarray}
By varying $\theta$ over the range $(0, \pi/2)$, accurate values of $\P(\X=\zero)$ would enable the recovery of the (integral) coefficients of the weight-enumerator polynomial of $Ker_L(P)$, which by choice of $P$ may be set to be any appropriately sized linear binary code we please.  The recovery of arbitrary weight-enumerator polynomials is $\mathbf{P^{GapP}}$-hard \cite{lit:vyalyi}.  \hfill \qed

\subsection{Background}

There has been a wide range of work into discovering restricted models of quantum computation which \emph{are} classically simulable.  For example, quantum circuits generating limited forms of entanglement, with classical simulations based on analysing matrix product states or contracting tensor networks; these circuits have a particularly constrained `circuit-topology', which leads to their simplicity (see \cite{lit:Mar05} for a summary of known results).  There is no particular circuit-topology imposed in our Z-network architecture (discussed in section~\ref{sect:architectures}), so it seems unlikely that the same methods would apply here.
Other positive classical simulability results include the stabiliser circuits of the Gottesman-Knill theorem and various matchgate constructions (see \cite{lit:Val02, lit:Jo08, lit:SWVC08, lit:SWVC2} and references therein).  These constructions differ significantly from our Z-networks in terms of the underlying algebra, the group generated by the set of allowable gates. 

H\o{}yer and Spalek \cite{lit:Hoy02} have shown that Shor's algorithm for Integer Factorisation can indeed be performed within a \emph{constant} number of timesteps on a Graph State processor (discussed in section~\ref{sect:architectures}), though their constructions offer no reason to believe that that constant might be smaller than, say, $\sim 100$; and of course, a general methodology for reducing the inherent time-complexity of oracle-dependent quantum search algorithms is known to be impossible, due to lower bounds on Grover's algorithm.  Dan Browne \cite{lit:Browne06} wrote about \emph{CD-decomposability}, which is the first rigorous treatment that we know of that explicitly links Graph State temporal depth with commutativity of Hamiltonian terms used to simulate a Graph state computation. 

Dan Simon \cite{lit:Si97} wrote about algorithms that use nothing more than an oracle and a Hadamard transform, and which therefore could be described as `temporally unstructured'.  However, his notion of `oracle' was one tailored for a universal quantum architecture, being essentially an arbitrarily complex general unitary transformation, and since there is no natural notion of one of these within our `temporally unstructured' paradigm, there is no real sense in which Simon's algorithm can count as an example of an algorithm within the $\mathbf{BPP^{IQP}}$ framework.  In particular, Simon's oracle implements a unitary that does \emph{not} commute with the Hadamard transform.

\subsection{Conjectures, implicit and explicit}

It is possible to form various hardness conjectures about the classical simulation of these $\IQP$ probability distributions.  
For a randomly chosen X-program $P$ of a given width $n$, it seems likely that the associated $\IQP$ distribution would be exponentially close to flat random.  Conditioned on its \emph{not} being random, there is no particular reason to think it would be approximately efficiently classically samplable.    
Here is an example of one such conjecture, though the precise details are not important to our arguments.

\medskip
\begin{conjecture}  \label{conj:sample}
There exists a distribution $\mathcal{D}$ on the set of X-programs, for which no classical Turing machine can gain a non-negligible $\Omega(1/poly)$ advantage in deciding whether or not the distribution associated to an X-program chosen randomly from $\mathcal{D}$ is exponentially close in trace distance to the uniform distribution.
\end{conjecture}
\medskip

This particular hardness conjecture is not quite what we really \emph{require}, but it gives an example of a plausible conjecture about classical simulation, and implies that for almost any X-program of interest, there is a certain \emph{event} (subset of output possibilities) whose probability will (probably) be estimated wrongly by your favourite classical polynomial-time event-probability-estimating device.
(Many similar conjectures sound equally plausible, in an area where almost nothing is known for sure.)
The point to emphasise in context of our two-player interactive protocol games is that it is not unreasonable for Alice to \emph{believe} that Bob can have no classical cheating strategy \emph{so long as} none such has been published nor proven to exist; and so our protocol may still serve as a demonstration (if not a proof) of a genuinely quantum computing phenomenon, despite the lack of proof of any simulation conjecture.

Another conjecture implicit in Alice's ability to make a fair hypothesis test---so that Bob will indeed have a good chance of passing the test when he does have an $\IQP$ oracle (or approximate version of one), but will stand little chance of faking a proof if relying on guesswork and (known) classical techniques---is one that ensures that Alice's X-programs really do incorporate a non-negligible amount of entropy.  Although we have little to go on besides scant simulation evidence from small examples, we want to make a conjecture that collision entropy is close to maximal within at least one relevant family of random constant-action X-programs.

\medskip
\begin{conjecture}  \label{conj:entropy}
The expected collision entropy of the probability distribution of a randomly selected X-program of width $n$, with constant action $\pi/8$, scales as $n - O(1)$ with the size of the program.
\end{conjecture}
\medskip

This conjecture is perhaps not directly relevant to the `hardness' of the $\IQP$ paradigm itself, but merely relevant to our game construction.
Note, for example, that since arbitrary---or random---obfuscation rows are used in the construction of the matrix $P$ in the construction of section~\ref{sect:protocol}, it follows that there will be much about the random variable $\X$ that is arbitrary---or `typical' in some vague sense---to the point that if one were sure that the only structure of significance were the hidden `causal' code $\Code_\s$, one could hope to approximate the distribution for $\X$, using knowledge of $\s$, by sampling uniformly at random (no biases) and applying a post-filter to create a bias in direction $\s$ of the required strength.  This gives some context for conjecture~\ref{conj:entropy}.

\subsection{Classical approximations}  \label{sect:classical}

Rather than speculate at this stage on which of the very many possible conjectures may or may not be true, we instead turn back to an examination of the mathematical structures underpinning the probability distributions in question.

Suppose we wish to construct a probability distribution that arises from some purely classical methods, which can be used to approximate our $\IQP$ distribution.  Our motivation here is to check whether any purported application for an $\mathbf{IQP}$ oracle might not be efficiently implemented without any quantum technology.  We proceed using the relatively \emph{ad hoc} methods of linear differential cryptanalysis.

\subsubsection{Directional derivatives}

For the case $\theta = \pi/8$, we will need to consider only second-order derivatives.  The same sort of method will apply to the case $\theta = \pi/2^{d+1}$ using $d$th order derivatives, but the presentation would not be improved by considering that general case here.

In terms of a binary matrix/X-program $P$, proceed by defining 
\begin{eqnarray}  \label{eqn:def_f}
  f     &:&  \F_2^n ~\rightarrow~ \Z/16\Z, \nonumber \\
  f(\a) &\equiv& \sum_{\p \in P} (-1)^{\p \cdot \a^T} \pmod{16},
\end{eqnarray}
and notate discrete directional derivatives as 
\begin{eqnarray} \label{eqn:deriv}
  f_\d(\a) &\equiv& f(\a) - f(\a\oplus\d) \pmod{16}.  
\end{eqnarray}

Consider also the \emph{second} derivatives of $f$, given by
\begin{eqnarray}  \label{eqn:deriv2}
  f_{\d,\e}(\a) 
    &\equiv&  f_\e(\a) ~-~ f_\e(\a\oplus\d) \pmod{16} \nonumber \\
    &\equiv&  2\sum_{\p \in P_\e} (-1)^{\p \cdot \a^T} 
              \left( 1 ~-~ (-1)^{\p \cdot \d^T} \right) \pmod{16} \nonumber \\
    &\equiv&  4\!\!\!\!\sum_{\p \in P_\d \cap P_\e} (-1)^{\p \cdot \a^T} \pmod{16} \nonumber \\
    &\equiv&  4\!\!\!\!\sum_{\p \in P_\d \cap P_\e} ~\prod_{j:p_j=1} \Bigl(~ 1 ~-~ 2a_j ~\Bigr) \pmod{16} \nonumber \\
    &\equiv&  \sum_{\p \in P_\d \cap P_\e} \left(~ 4 ~+~ 8\!\!\!\sum_{j~:~p_j=1} \!\!a_j ~\right) \pmod{16},
\end{eqnarray}
each of which is quite patently a linear function in the bits $(a_1, \ldots, a_n)$ of $\a$, as a function with codomain the ring $\Z/16\Z$, regardless of the choice of directions $\d,\e$.

\medskip
\begin{lemma}
With $f$ defined as per line~(\ref{eqn:def_f}), and $\X$ the random variable of lemma~\ref{lem:thing}, for all~$\s$,
\begin{eqnarray}  \label{eqn:piby8}
  \P( \X \cdot \s^T = 0 ) 
    &=&  \E_\a \left[ \cos^2\Bigl(~ \frac\pi8 \cdot f_\s(\a) ~\Bigr) \right],
\end{eqnarray}
and so the $\IQP$ probability distribution (in the case $\theta=\pi/8$) may be viewed as a function of $f$ rather than as a function of $P$.
\end{lemma}
\medskip

\proof
Starting from the proof of theorem~\ref{thm:bias}, line~(\ref{eqn:wooo}), 
\begin{eqnarray}
  \P(\X \cdot \s^T = 0)
    &=& \frac12\left(~ 1 ~+~ \E_\a \left[ e^{ i\theta \sum_\p (-1)^{\p \cdot \a^T} \Bigl(1 - (-1)^{\p \cdot \s^T}\Bigr) } \right]
        ~\right) \nonumber \\
    &=& \frac12\left(~ 1 ~+~ \E_\a \left[ \exp\left( \frac{i\pi}8 \bigl( f(\a) - f(\a\oplus\s) \bigr) \right) \right] ~\right) \nonumber \\
    &=& \frac12\left(~ 1 ~+~ \E_\a \left[ \cos\Bigl(~ \frac\pi8 \cdot f_\s(\a) ~\Bigr) \right] ~\right).
\end{eqnarray}
The second line above is obtained immediately from the first, using the definition of $f$.  
The third line follows because the expression is real-valued.  
The conclusion follows from a basic trigonometric identity, and linearity of the expectation operator.
\hfill \qed \medskip

And so \emph{if} there is a hidden $\s$ such that $\P( \X \cdot \s^T = 0 )=1$, \emph{then} that implies $f_\s(\a) \equiv 0 \pmod{16}$ for all $\a$.
This is essentially a non-oracular form of the kind of function that arises in applications of Simon's Algorithm \cite{lit:Si97}, with $\s$ playing the role of a \emph{hidden shift}.
One could find linear equations for such an $\s$ if it exists, because it would follow immediately that $f_{\d,\e}(\s) = f_{\d,\e}(\zero)$ for any directions $\d,\e$, which is by line~(\ref{eqn:deriv}) equivalent with 
\begin{eqnarray}
  \left( \sum_{\p \in P_\d \cap P_\e} \!\!\p \right) \cdot \s^T &=& 0.
\end{eqnarray}

\subsubsection{Classical sampling}

To make use of this specific second-order differential property, we need to analyse the probability distribution that a classical player can generate efficiently from it.
Proceed by defining a new probability distribution for a new random variable $\Y$, as follows~:
\begin{eqnarray}  \label{eqn:classprob}
  \P(\Y=\y) 
    &=&  \P_{\d,\e} \left(~ \sum_{\p \in P_\d \cap P_\e} \!\!\p ~=~ \y ~\right).
\end{eqnarray}
This may be classically rendered, simply by choosing $\d,\e \in \F_2^n$ independently with a uniform distribution, and then returning the sum of all rows in $P$ that are not orthogonal to either $\d$ or~$\e$.

\medskip
\begin{lemma}  \label{lem:classprob}
The classical simulable distribution on the random variable $\Y$ defined in line~(\ref{eqn:classprob}) satisfies
\begin{eqnarray}  \label{eqn:classfromcode}
  \P( \Y \cdot \s^T = 0 ) 
    &=&  \P\Bigl(~ \c_1^T \cdot \c_2 = 0 ~~|~~ \c_1,\c_2 \sim \Code_\s ~\Bigr) \\
    &=&  \frac12\left(~ 1 ~+~ 2^{-rank(~ P_\s^T \cdot~ P_\s ~)} ~\right),
\end{eqnarray}
and so the bias of $\Y$ in direction $\s$ is a function of the matroid $P_\s$.
\end{lemma}
\medskip

\proof
Starting from line~(\ref{eqn:classprob}),
\begin{eqnarray}
  \P( \Y \cdot \s^T = 0 )
    &=& \sum_{\y ~:~ \y \cdot \s^T = 0}  ~~ \P_{\d,\e} \left(~ \sum_{\p \in P_\d \cap P_\e} \!\!\p ~=~ \y ~\right) \\
    &=& \P_{\d, \e} \left(~ \sum_{\p \in P_\d \cap P_\e} \!\!\p \cdot \s^T ~=~ 0 ~\right) \nonumber \\
    &=& \P_{\d, \e} \left(~ wt(~ P\cdot\d^T ~\wedge~ P\cdot\e^T ~\wedge~ P\cdot\s^T ~) \equiv 0 \pmod2 ~\right) \nonumber \\
    &=& \P_{\d, \e} \left(~ wt(~ P_\s\cdot\d^T ~\wedge~ P_\s\cdot\e^T ~) \equiv 0 \pmod2 ~\right) \nonumber \\
    &=& \P_{\d, \e} \left(~ \d\cdot P_\s^T \cdot P_\s\cdot\e^T = 0 ~\right). \nonumber
\end{eqnarray}
The \emph{wedge operator} $\wedge$ here denotes the logical \emph{AND} between binary column-vectors.

The first line of the lemma follows from the obvious substitutions $\c_1 = P_\s \cdot \d^T$, $\c_2 = P_\s \cdot \e^T$.
The second line follows because unimodular actions on the left or right of a quadratic form (such as $(P_\s^T \cdot P_\s)$) affect neither its rank nor the probabilities derived from it; so it suffices to consider the cases where it is in Smith Normal Form, \emph{i.e.} diagonal, which are trivially verified.    
Since this expression is patently invariant under invertible linear action on the right and permutation action on the left of $P_\s$, it too is a matroid invariant.  
\hfill \qed

\subsubsection{Inter-relation}

Thus we have established some kind of correlation between random variables $\X$ and $\Y$.
\begin{theorem}  \label{thm:unitbound}
In the established notation, for X-programs with fixed $\theta=\pi/8$,
\begin{eqnarray}  \label{eqn:unitbound}
  \P( \X \cdot \s^T = 0 ) = 1 
    ~~&\Rightarrow&~~  \P( \Y \cdot \s^T = 0 ) = 1.
\end{eqnarray}
\end{theorem}  

\proof
By theorem~\ref{thm:bias}, the antecedent gives, for all $\c \in \Code_\s, ~n_\s \equiv 2 wt(\c) \pmod8$, where $n_\s$ is again the length of the code $\Code_\s$.
This entails that every codeword in $\Code_\s$ has the same weight modulo 4, including the null codeword, so $\Code_\s$ must be doubly even\footnotemark{}.
\footnotetext{\emph{Doubly even} just means that every codeword has weight a multiple of 4.}
It is easy to see that doubly even linear codes are self-dual.\footnotemark{}
\footnotetext{\emph{Self-duality} just means that the dual code---consisting of all words orthogonal to every codeword---is equal to the code.}
%of $\Code$ just means that $\{ \c' \in \F_2^n ~:~ \forall \c \in \Code, ~ \c^T \cdot \c' = 0 \}  = \Code$.} 
%So if $\c_1$ and $\c_2$ are codewords, so is $\c_1 + \c_2$.  Note that $wt(\c_1 + \c_2) \equiv wt(\c_1) + wt(\c_2) + 2\c_1^T \cdot \c_2 \pmod4$, and so $\c_1$ and $\c_2$ are always orthogonal~: the code is self-dual.  
By lemma~\ref{lem:classprob}, the consequent is obtained.
\hfill \qed \medskip

The only counterexamples to the \emph{converse} implication seem to occur in the trivial cases whereby the binary matroid $P_\s$ has circuits of length 2, \emph{i.e.} where $P_\s$ has repeated rows. 

Note that if $\P( \X \cdot \s^T = 0 )$ were equal to 1 precisely, then by making a list of samples from $\IQP$ runs, storing them in a matrix, and performing Gaussian Elimination to recover the kernel of the matrix, it would be straightforward to compute the hidden $\s$.  However, theorem~\ref{thm:unitbound} shows that this is exactly the condition required for being able to compute $\s$ via purely classical means.  For this reason, it seems hard to find decision languages that plausibly lie in $\mathbf{BPP^{IQP}} \backslash \mathbf{BPP}$.

This random variable $\Y$ is the `best classical approximation' that we have been able to find for $\X$.  
(The intuition is that it captures all of the `local' information in the function $f$, which is to say all the `local' information in the matroid $P$, so that the only data left unaccounted for and excluded from use within building this classical distribution is the `non-local' matroid information, which is readily available to the quantum distribution via the magic of quantum superposition.) 
There seems to be no other sensible way of processing $P$ (or $f$) classically, to obtain useful samples efficiently, though it also seems hard to make any rigorous statement to that effect.

\medskip
\begin{conjecture}  \label{conj:best}
The classical method defined in this section, yielding random variable $\Y$, is asymptotically classically optimal (when comparing worst-case behaviour and restricting to polynomial time) for the simulation of $\IQP$ distributions arising from constant-action $\theta=\pi/8$ X-programs.
\end{conjecture}
\medskip

This conjecture lends credence to the design methodology of section~\ref{sect:protocol}.

\subsection{Future work}

We might also recommend the further study of matroid invariants through quantum techniques, or perhaps the invariants of \emph{weighted} matroids, since they seem to be the natural objects of $\mathbf{IQP}$ computation as hitherto circumscribed.
This would seem to be fertile ground for developing examples of things that only genuine quantum computers can achieve.

Note that if it weren't for the correlation described in theorem~\ref{thm:unitbound}, then it would be possible to conceive of a mechanism whereby an $\mathbf{IQP}$-capable device could compute an actual secret or witness to something (\emph{e.g.} learn $\s$), so that the computation wouldn't require two rounds of player interaction to achieve something non-trivial.  Yet as it stands, it is an open problem to suggest tasks for this paradigm involving no communication nor multi-party concepts.

\section{Architectures}  \label{sect:architectures}

In this section we sketch two more architectures for implementing $\IQP$ oracles.  These architectures are probably more physically feasible than X-programs.

\subsection{Z-networks}

The network- or circuit-model of computation is perhaps the most familiar one.  
Programs for the first architecture we call ``Z-networks'', since the program is most easily described as a network of gates on an array of qubits, where the allowed gate-set includes just the Controlled-Not gate from any qubit to any qubit and the single-qubit gate that implements the Pauli $Z$ Hamiltonian\footnotemark{} for some time.
\footnotetext{$Z = \ketbra00 - \ketbra11$, on a single qubit.}  
Although this Z-network architecture \emph{does} have a notion of temporal structure---because it is important the order in which the gates of the network are carried out---nonetheless it is useful for our analysis because it turns out to have effectively the same computational power as the X-program architecture under some basic assumptions, and the Lie group structure underpinning the kinds of transformation allowable within the Z-network architecture is particularly easy to work with.

On the understanding that $n$ qubits are initialised into $\ket{\zero}$ in the computational basis and ultimately measured in the computational basis, it is well known that the gate-set consisting of Controlled-Not gates together with \emph{all single-qubit rotations} is universal for \textbf{BQP}, and the Lie group generated by this gate-set generates the whole of $SU(2^n)$, modulo global phase.
However, by the term ``Z-network'' we mean explicitly to limit \emph{the single-qubit gates} to being those which implement $e^{i \theta Z}$ for some action $\theta$, so that the Lie group spanned by the gate-set (represented in the computational basis) consists of unitary matrices that are supported by permutation matrices, \emph{i.e.} those unitaries that have just one non-zero entry per row.  Any such unitary can be factored into a diagonal matrix followed by a permutation matrix.  (In all cases, global phase is to be regarded as physically irrelevant, and may be `quotiented out' from the groups in question.)

We can describe groups by giving generator sets for them.  
The group containing \emph{all even} permutations and all diagonal elements (modulo global phase) is
\begin{eqnarray}  \label{eqn:FullLiegroup}
  \left<~ \mbox{Toffoli, C-Not, } X, ~e^{i\theta Z} ~\right>  &=&  \left<~ \mbox{any even permutation}, ~\mbox{any diagonal} ~\right>.
\end{eqnarray}
This `qualifies' as a Z-network group; indeed, all Z-network groups are to be a subgroup of this one.  But for the purposes of comparison with X-programs and the $\IQP$ paradigm, it suffices to consider the much simpler group given by
\begin{eqnarray}  \label{eqn:Liegroup}
  \left<~ \mbox{C-Not, } X, ~e^{i\theta Z} ~\right>  &=&  \left<~ \mbox{any linear permutation}, ~\mbox{any diagonal} ~\right>.
\end{eqnarray}
This latter group does not apparently contain (efficiently) the dynamics of classical computation.
(The $X$ gate is necessary to enable the construction of all diagonals, but one might prefer to replace implementation of $X$ by the availability of an ancilla $\ket1$ qubit, so that a C-Not gate can simulate an $X$ gate.  In the language of complexity theory, (\ref{eqn:FullLiegroup}) might be said to stand in relation to~(\ref{eqn:Liegroup}) as $\mathbf{P}$ stands to $\mathbf{\bigoplus L}$.)

One can see how to build a variety of constructions within the group at line~(\ref{eqn:Liegroup}), using the specified gate-set. \emph{E.g.} by conjugating $e^{i\theta Z_1}$ with two C-Not gates one can create an $e^{i\theta Z_1 Z_2}$ composite unitary, and similarly $e^{i\theta Z_1 Z_2 Z_3}$ can be created, \emph{etc}.

\subsubsection{Reductions between Z-networks and X-programs}

This neat mathematical structure (\emph{i.e.} as at line~(\ref{eqn:Liegroup})) enables probability distributions of the kind at line~(\ref{eqn:dist1}) to be simulated.

\medskip
\begin{lemma}
A Z-network can always be designed to simulate an X-program efficiently.
\end{lemma}
\medskip

\proof
Simply by initialising the input qubits to the Z-network to $\ket{+^n}$ in the Hadamard basis, and measuring output in the same Hadamard basis, the simulation of a given X-program, $P$, proceeds by constructing a small network of CNot gates and an $\exp(i \theta_\p Z)$ gate for each row $\p$ of $P$.  Details omitted for brevity.  \hfill \qed

Conversely, 
\begin{lemma}
An X-program can efficiently simulate a given Z-network, provided that that Z-network uses Hadamard-basis input, C-Nots, $X$ gates, and $e^{i\theta Z}$ gates only, and outputs in the Hadamard basis.  
\end{lemma}
\medskip

\proof
The required reduction just associates one X-program element $(\theta_\p, \p)$ to each $e^{i\theta Z}$ gate, setting $\theta_\p \leftarrow \theta$ and specifying $\p$ according to the location of the $e^{i\theta Z}$ gate \emph{and} the totality of C-Not gates to the left of that gate.  A final piece of simple post-processing is needed after the measurement phase of the X-program, to account for the C-Nots in the Z-network, but this post-processing simply consists in applying the same C-Nots (with directions reversed) on the classical measurement outcomes.  (This is because moving from the Hadamard basis to the computational basis has the effect of reversing the direction of C-Not gates.)  
\hfill \qed \medskip

The point of these reductions is to highlight the sense in which the group associated to simple Z-networks stands in the same relation to  the set of X-programs as the `full' $SU(2^n)$ Lie group stands to the set of proper full-blown quantum algorithms.

\subsection{Graph-programs}

Programs for the second of these architectures we call ``Graph-programs'', since the program is most easily described as the construction of a graph state followed by a series of measurements of the qubits in the graph state in various bases \cite{lit:Browne06}. 
Graph state computing architectures are popular candidates for scalable fully universal quantum processors \cite{lit:Raus01, lit:Raus03}.  
Here we are concerned not with universal architectures, but with the appropriate restriction to `unit time' computation.  So, unlike universal graph state computation, our Graph-programs do not admit adaptive feed-forward, which is to say that all measurement angles must be known and fixed at compile-time, so that all measurements can be made simultaneously once the graph state has been built.  In this sense, the `depth' of a Graph-program is 1.
A graph state has qubits that are initially devoid of information, but which are entangled together according to the pattern of some pre-specified graph.
A graph state can be constructed without inherent temporal complexity, perhaps even prepared in a single computational time-step, because there is no implicit reason requiring one edge of the graph to be prepared before any other.
(It is still fair to argue that the circuit-depth of the process that generates a graph state is linear in the valency of the graph, but that is not a measure of `inherent' temporal complexity.)  
We will show how Graph-programs can simulate the output of X-programs if a little trivial classical post-processing of the measurement results is allowed.

A Graph-program is taken to be an undirected (usually bipartite) graph with labelled and distinguished vertices.  
The vertex set is denoted $V$, of cardinality $n$, and for each $v \in V$ there is an element of $SU(2)$ labelling it; $R_v \in SU(2)$.  
The edge set is denoted $E$.
To implement the program, a qubit is associated with each vertex and is initialised to the state $\ket+$ in the Hadamard basis.  
Then a Controlled-$Z$ Pauli gate is applied between each pair of qubits whose vertices are a pair in $E$.  Since these Controlled-$Z$ gates commute, they may be applied simultaneously, at least in theory.
Finally, each vertex qubit $v$ is measured in the direction prescribed by its label $R_v$, returning a single classical bit.  Clearly the order of measurement doesn't matter, because the measurement direction is \emph{prescribed} rather than \emph{adaptive}.
Hence a sample from $\F_2^n$ (a bit-string) is thus generated as the total measurement result.

Combining this together, we see that the probability distribution for such an output is 
\begin{eqnarray}  \label{eqn:dist2}
  \P(\X=\x) 
  &=& \left| \bra\x ~\prod_{v \in V} R_v 
                 \cdot \!\!\!\!\prod_{(u,v) \in E}\!\!\!\frac{1 + Z_u + Z_v - Z_u Z_v}2 
                ~ \ket{+^n} \right|^2.  
\end{eqnarray}
Here the measurement has been written using the notation of the computational basis, with an appropriate (passive) rotation immediately prior.

\subsubsection{Graph-programs and X-programs}

Having described three architectures, we've indicated that the X-programs, characterised by the formulation at line~(\ref{eqn:dist1}), are in some natural sense the `lowest common denominator' amongst the architectures (and unitary groups) of interest. 

\medskip
\begin{lemma}
A Graph-program can always be designed to simulate an X-program efficiently.
\end{lemma}
\medskip

\proof
Suppose we're given an X-program, written $\{~ (\theta_\p, \p) ~:~ \p \in P \subset \F_2^n ~\}$.  Then it is straightforward to simulate it on a Graph State architecture, as follows.  
Let $V$ be the disjoint union of $[1..n]$ and $P$, so that the graph state used to simulate the program will have one \emph{primal} qubit for each qubit being simulated, plus one \emph{ancilla} qubit for each program element $\p$, the total cardinality of $V$ being polynomial in $n$, by hypothesis.
Let $(j,\p) \in E$ exactly when the $j$th component of $\p$ is a 1.  In this way, the resulting graph is bipartite, linking primal qubits to those ancill\ae{} that they have to do with.  Let $R_j$ be the Hadamard element ($H$) for all primal qubits, so that all primal qubits are measured in the Hadamard basis.  Let $R_\p = \exp( i\theta_\p X )$, so that every ancilla qubit is measured in the $(YZ)$-plane at an angle specified by the corresponding program element.

If the resulting Graph-program is executed, it will return a sample vector $\x \in \F_2^{n+\#P}$ for which the $n$ bits from the primal qubits are correlated with the $\#P$ bits from the ancill\ae{} in a fashion which captures the desired output, (though these two sets separately---marginally---will look like flat random data.)  To recover a sample from the desired distribution, we simply apply a classical Controlled-Not gate from each ancilla bit to each neighbouring primal bit, according to $E$, and then discard all the ancilla bits.
One can use simple circuit identities to check that this produces the correct distribution of line~(\ref{eqn:dist1}) precisely.  \hfill \qed \medskip

\subsection{Physical comparison}

An X-program uses only `one timeslice', but presumes an arbitrarily large gatespan (interaction length), up to $n$, the number of qubits in the X-program.

The Z-network reduction is physically preferable because it uses small gates (gatespan = 2) to simulate an X-program.  However, it has network depth that is not constant; rather, it is likely to be quadratic in $n$, in general, unless one is careful to make optimisations when `compiling' the network.  

The Graph-program reduction uses more qubits than $n$, but has better depth properties than the Z-network architecture.  Again, only small gates are used.
It seems natural to suggest that the `true temporal cost' of implementing a Graph program would either be one process for measurement plus either one process or $degree$ processes for initialising the graph state, where $degree$ denotes the vertex-degree of the underlying graph.  
The number of qubits required in the simulation of an X-program is also slightly larger, since ancilla qubits are used.
The worst case in our interactive protocol application would therefore require about $5q/2$ qubits, rather than $q/2$, though the vertex-degree would no bigger than the number of qubits.  
The kinds of graph called for in our Graph-program reduction are not the usual cluster state graphs (regular planar lattice arrangement) used in measurement-based quantum computation. 
The bipartite graphs described in the reduction will usually be far from planar for generic X-programs, having a relatively high genus.  This means that the graph cannot be `laid out' on a plane without edges crossing.

%% ACKNOWLEDGEMENTS

\section*{Acknowledgements}

We would like to thank Tobias Osborne, Richard Jozsa, Ashley Montanaro, Dan Browne, Scott Aaronson, and Richard Low for useful discussions and suggestions.  We also acknowledge the support of the EC-FP6-STREP network QICS.

%% APPENDIX

\section*{Appendix}

A proof of theorem~\ref{thm:bias}. 
Throughout, the variable $\p$ ranges over the rows of the binary matrix $P$, which are the program elements of an X-program. 

Derive line~(\ref{eqn:walshcode}) from line~(\ref{eqn:dist1}) in the case that the value $\theta$ is constant.

\begin{eqnarray} 
  \P(\X=\x)
  &=& \left| \bra\x ~\exp\left(~\sum_\p i\theta_\p \bigotimes_{j:p_j=1} X_j~\right)~ \ket{\zero^n} \right|^2  \nonumber \\
  &=& \left| 2^{-n}\sum_\a (-1)^{\x \cdot \a^T}\bra\a ~~\exp\left(~\sum_\p i\theta_\p \bigotimes_{j:p_j=1} Z_j~\right)~~ 
                   \sum_\b \ket{\b} \right|^2 \nonumber \\
  &=& \left| ~\E_{\a}~ \left[ (-1)^{\x \cdot \a^T} ~\exp\left(~i\theta \sum_{\p} (-1)^{\p \cdot \a^T} ~\right) 
                              \right] ~\right|^2 \nonumber \\
  &=& \E_{\a,\d}~ \left[ (-1)^{\x \cdot \d^T} ~\exp\left(~ i\theta \sum_{\p} (-1)^{\p \cdot \a^T} 
                                                           \Bigl(1 - (-1)^{\p \cdot \d^T}\Bigr) ~\right) \right].  
\end{eqnarray}
On the second line we made a change of basis, so as to replace the Pauli $X$ operators with Pauli $Z$ ones.

\begin{eqnarray}  \label{eqn:wooo}
  \P(\X \cdot \s^T = 0)
    &=& 2^{n} ~\E_{\x} \left[~ \{\x\cdot\s^T=0\} \cdot \P( \X = \x ) ~\right] \nonumber \\
    &=& 2^{n} ~\E_{\a,\d,\x} 
        \left[ \frac{(1+(-1)^{\x\cdot\s^T})}2 ~(-1)^{\x \cdot \d^T}
        ~e^{ i\theta \sum_\p (-1)^{\p \cdot \a^T} \Bigl(1 - (-1)^{\p \cdot \d^T}\Bigr) } \right] \nonumber \\
    &=& 2^{n} ~\E_{\a,\d} 
        \left[ \frac{\Bigl( \{\d=\zero\} + \{\d=\s\} \Bigr)}2 
             ~e^{ i\theta \sum_\p (-1)^{\p \cdot \a^T} \Bigl(1 - (-1)^{\p \cdot \d^T}\Bigr) } \right] \nonumber \\
    &=& \frac12\left( 1 ~+~ \E_\a \left[ e^{ i\theta \sum_\p (-1)^{\p \cdot \a^T} \Bigl(1 - (-1)^{\p \cdot \s^T}\Bigr) } \right]
        \right).
\end{eqnarray}
These transformations are conceptually simple but notationally untidy.
%The last line above can be used in a classical simulation of any \emph{very small} (up to $\sim20$ qubits) $\mathbf{IQP}$ process.

\begin{eqnarray}
  2 \cdot \P(\X \cdot \s^T = 0) - 1 
      &=& \sum_j e^{ij\theta} ~\E_{\a,\phi} 
        \left[ e^{i\phi\left( -j ~+~ \sum_\p (-1)^{\p \cdot \a^T} \Bigl(1 - (-1)^{\p \cdot \s^T}\Bigr) ~\right)} \right] \nonumber \\
      &=& \sum_j e^{ij\theta} ~\P_{\a} 
        \left(~ j ~=~ 2\!\!\!\!\!\!\sum_{\p~:~\p\cdot\s^T=1} \!\!(-1)^{\p \cdot \a^T} ~\right)  \nonumber \\
      &=& \sum_j e^{ij\theta} ~\P 
        \left(~ j = 2 (~ n_\s - 2 \cdot wt( \c ) ~) ~~|~~ \c \sim \Code_\s ~\right) \nonumber \\
      &=& \sum_w \cos(~ 2\theta(n_\s - 2 w) ~) \cdot \P\left(~ w = wt( \c ) ~~|~~ \c \sim \Code_\s ~\right).
\end{eqnarray}
Here we have used the standard Fourier decomposition of a periodic function, and used the fact that the function is known to be real.  The variable substitution at the third line was $\c = P_\s \cdot \a^T$, understood in the correct basis.  At the fourth line it was $w = (2n_\s-j)/4$.

\begin{eqnarray*}
  \P(\X \cdot \s^T = 0)
    &=& \sum_{w=0}^{n_\s} \cos^2(~ \theta(n_\s - 2 w) ~) \cdot \P\left(~ w = wt( \c ) ~~|~~ \c \sim \Code_\s ~\right) \\ 
    &=& \E_{\c \sim \Code_\s} \left[~ \cos^2\Bigl(~ \theta( n_\s ~-~ 2 \cdot wt(\c) ) ~\Bigr) ~\right]. ~~~~~~~~~~~~~~~~~~~~~~~ \qed
\end{eqnarray*}

%% REFERENCES
%\vbox{

%}%% Kill the vbox if you add more refs.

\end{document}